# Tourism diversification paths in ski mid-mountain territories: any transformations?

**Laura Rouch**
*Univ. Grenoble Alpes, INRAE, LESSEM, Grenoble, France*

**Emmanuelle George**
*Univ. Grenoble Alpes, INRAE, LESSEM, Grenoble, France*

**Abstract**
In the context of adapting to global changes, the diversification of the tourist offer seems to be a solution for the mid-mountain areas, which are structured around and dependent on ski tourism. However, it remains unclear how tourism diversification occurs, what forms it can take and what it produces at a territorial scale. In this study, we apply a theory of regional diversification to the case of tourism in order to examine tourism diversification paths. We base our analysis on qualitative data collected on two French study areas: the intercommunities of the massif du Sancy and of the Haut-Chablais. Our results show a pattern in tourism diversification paths, following three steps that can lead to the abandonment of the perimeter of the ski resort. However, among the different types of tourism diversification trajectories, only a saltation form of tourism diversification can lead to the establishment of a larger and more diversified tourism system. Our findings also show that several types of tourism diversification paths can coexist at a territorial scale.

**Introduction**
Since the 1990s, in France, mid-mountain touristic areas have been grappling with their future (Pascal, 1993). These regions, characterized by altitude, slope, and mosaic landscapes of meadows and forests, as well as limited access to high-altitude areas and interdependencies with lower-lying regions, serve as complex spatial systems where public interventions reflect socio-economic fragility (Rieutort, 1997; Diry, 1995). Inherently vulnerable, these mid-mountain regions are mostly specialized in ski tourism, and are now facing the profound impacts of climate change on snow reliability (Steiger et al., 2019). Amidst these challenges, strategies such as snowmaking and alternative tourism ventures have arisen (Burki et al., 2003). French public policy since the 1990s has focused on diversifying tourism offerings in response (Achin, 2015; François, 2007), first aiming to adapt to structural changes – such as changes in tourist demand behavior or in snow reliability – and now also to experiment with ways of transitioning specialized mountain economies (George & Achin, 2020).

In the literature, the term "tourism diversification" has gained prominence since the early 2000s, encompassing various processes across different scales (Rouch, 2022; Weidenfeld, 2018). While presented as a strategy to adapt to climate change (Abegg et al., 2007; Burki et al., 2003) or as a territorial process (François et al., 2013; George & Achin, 2020; Perret, 1992) in our areas of interest, tourism diversification is also conceptualized as a means to rejuvenate a tourist destination (Farmaki, 2012) or to adapt a tourism product (Benur & Bramwell, 2015; Erkuş-Öztürk & Terhorst, 2015). These studies emphasize the identification of steps in tourism diversification and the exploration of its spatial and temporal dimensions. Although not exclusively confined within this theoretical realm, these works underscore the relevance of an evolutionary approach (Boschma & Martin, 2007; Weidenfeld, 2018). Following the Evolutionary Economic Geography (EEG) field on tourism (Ma & Hassink, 2013; Brouder et al., 2016) – and yet being open to other theoretical frameworks on territorial evolution – would allow us to examine and qualify the evolutionary drivers and trajectories associated with tourism diversification in place-based economies. Yet, within this literature, few have thoroughly explored the intricate and long-term aspects of tourism diversification (Achin, 2015; Erkuş-Öztürk & Terhorst, 2015), and a gap remains in comprehending the evolutions and potential impacts associated with tourism diversification at a territorial-scale. In this context of French mid-mountain tourism territories facing environmental constraints, there is therefore limited explanation of how their economies, structured around ski tourism, have or have not transformed as a result of tourism diversification, and in what direction, even though they have been supported by public policies for 30 years now.

Consequently, this article seeks to address both theoretical and empirical objectives and gaps: offering an evolutionary perspective on tourism diversification, while empirically examining the ensuing territorial transformation. To achieve this, we propose a study frame to characterize tourism diversification as a trajectory, following Boschma et al. (2017) analysis framework and using qualitative data integrated in a chronosystemic timeline (Bruley et al., 2021). We will base our analysis on the two perimeters of the intercommunities of the massif du Sancy and the Haut-Chablais. These two sites, located in the same Auvergne-Rhône-Alpes administrative region, are examples of French mid-mountain tourist areas, structured around the snow economy and experimenting with diversification strategies in different ways in terms of content and timing.

**Conceptual background**

*A review of tourism diversification: purposes, scales and processes in the context of changing touristic mid-mountain territories*

Before delving into its implications, it is necessary to review the concept of tourism diversification and its applications. This review aims to clarify our definition within the context of evolving touristic mid-mountain territories. To date, only a few studies have interpreted ~~understood~~ tourism diversification as a trajectory and attempted to analyze its contributions in the transition and adaptation of ski tourism areas. It has been seen as a means to enhance ski areas performance (Botti et al., 2012) or as the evolution and structuring of the outdoor sports sector (Langenbach & Jaccard, 2019; Perrin-Malterre, 2018). Examining tourism diversification in such areas has also facilitated the identification of regionalization and territorial anchoring of economic activities, especially tourism, in the mountains (François, 2007). The concept of Localized Tourism Systems (LTS) (Capone, 2006; Perret, 1992) sanctifies this grounding and introduce tourism diversification as a way to question the relationship between the ski sector and the territory where it takes place (Marcelpoil, 2008). Align with the development of LTS, diversification is presented as a way to enhance the value of specific local resources (François, 2007), then examined according to the governance modalities that facilitate its implementation (Achin, 2015), all in the pursuit of adapting to global changes. This foundation in French literature enables the comprehension of tourism diversification as a regional process in specific touristic and vulnerable mountainous areas. However, it does not expound on the process itself, primarily focusing on its implementation and how it challenges specialization within each ski resort.

To acquire a more relevant understanding of the tourism diversification process itself, we have extracted insights from works covering various levels of tourism diversification. At the sectoral level, tourism diversification serves to reflect changes within the tourism sector (Bramwell, 2004) and to describe the expansion of often rural economies through tourism (Hjalager, 1996). Then, other works have attempted to delve into the stages, origins, and implications of tourism diversification, considering shifts within established tourism sectors and touristic destinations. Initially, tourism diversification was empirically scrutinized as a rejuvenation stage (Butler 1980; 2006) for declining sun, sea and sand destinations. This was achieved through the creation of distinct alternative offers, such as cultural tourism or agritourism (Farmaki, 2012; Sharpley & Vass, 2006). In this context, tourism diversification is summarized as a product differentiation at a certain point in time, rather than a continuous process. To answer this gap then, Erkuş-Öztürk & Terhorst, (2015) delve deeper and study the process of economic diversification in single-asset tourism destinations, accompanied with Benur & Bramwell, (2015) who develop a theoretical framework to identify the dimensions and steps of tourism product diversification. Tourism diversification is characterized as a series of actions initiated by tourism operators, such as promotion, creation of new tourism products, or adaptation of existing ones. In all these works, and whatever the scale, tourism diversification is seen as a way to continue growth. While these perspectives help uncover the origins of diversification and provide a temporal dimension, they fall short in considering tourism diversification as a regional – or a territorial (Weidenfeld, 2018) – evolution in a context of environmental constraints. In addition, none of these works apply to the mountainous context, known for its vulnerability.

Interested in the processual approach of tourism diversification – to deeply understand its ingredient and steps – and coupled with a more specific research on how mid-mountain territories do adapt, we mix the two types of works. This emphasis rests on a territorial scale (Weidenfeld, 2018) of tourism diversification in ski-tourism areas. Bridging the scales of the product and the sector, the regional or territorial scale of tourism diversification holds significance in contemplating its multi-scale nature, according to Weidenfeld (2018). At this scale, tourism diversification is characterized by forms of anchoring to supporting areas, leveraging local attributes and skills (Achin, 2015). More specifically, at the level of mid-mountain territories, we understand tourism diversification as 'a process leading to the creation of non-skiing tourism activities across all seasons; the integration of tourism initiatives from other economic sectors into the local tourism sector; and based on the development of new skills and competences beyond ski tourism' (Rouch, 2022). This process is driven by a variety of public, private, local and supra-local actors, including tourism operators and other economic actors proposing touristic activities located within the territory, local administrations and public policy makers. We consider tourism diversification at the territorial scale as an aggregation of local actors diversifying their economic activities – for instance a local administration opening a new museum, a farmer starting to offer farm visits to tourists and locals, or a ski rental company that now also hires bicycles for all seasons. Although we focus on the tourism purpose of local activities, we remain aware of the more global economic diversification it can represent at the scale of an actor and a territory. With this definition in mind, considering regional development and territorial anchoring dimensions, tourism diversification assumes an evolutionary perspective (Weidenfeld, 2018).

***Diverse forms of tourism diversification trajectories: a blend of Evolutionary Economic Geography and Transition Studies***

Examining and scrutinizing tourism diversification within mountain territories specialized in ski tourism raises questions about their evolutionary paths away from specialization. Similar to what has been demonstrated for other types of touristic regions, tourism diversification is a process capable of shaping the evolution trajectory of a destination (Erkuş-Öztürk & Terhorst, 2015). As previously noted by Weidenfeld (2018), tourism diversification can effectively be approached within the EEG realm. According to EEG, diversification, alongside specialization, is an explanatory variable for the difference in dependence on a specific economy between two regions (Boschma & Martin, 2007). Viewing touristic destinations as specific economic regions has led to the application of EEG principles to tourism, as first theorized by Ma & Hassink (2013) and Brouder and Ericksen (2013), confirmed by Brouder et al. (2016) and resulting in a growing body of literature (Sanz-Ibañez & Clavé, 2014; Clavé & Wilson, 2017; Gill & Williams, 2014; Ioannides et al., 2015; Niewiadomski & Brouder, 2022). Most EEG approaches in the context of tourism are theoretical and empirical. They analyze the potential pathways a touristic destination might follow and the underlying reasons, whether it involves path dependency (Brouder et al., 2016; Sanz-Ibañez & Clavé, 2014), path creation (Gill, 2018), path plasticity (Halkier & Therkelsen, 2013; Clavé & Wilson, 2017) or co-evolution (Ma & Hassink, 2013. Although EEG portrays regional diversification as a possible process out of dependency and gives a framework to analyze the different forms of diversification trajectories (Boschma et al., 2017), its application to tourism has yet to be comprehensively addressed and empirically detailed. Furthermore, the works discussing path dependency within touristic destinations do describe outcomes resembling what we perceive as tourism diversification – a result of departing from dependency. However, they don't fully capture it as a process or a trajectory. Returning to the French literature on the evolution of ski resorts, particularly the contributions of Marcelpoil (2008), inspired by Butler (1980, 2006) and following Perret (1992) LTS frameworks, tourism diversification is depicted as a possible path towards a more diversified tourism economy in the mountains. Consequently, we believe that tourism diversification can also be considered as a trajectory in the way it can explain the mechanism of change, the transitions from one form of network to another, or from one state of development to another (Chabault, 2009). Moreover, scrutinizing tourism diversification through the lens of a trajectory would allow us to specify the degree of transition of the mid-mountain territories, as to explain its origin and purpose.

To understand how tourism diversification transforms mountain economies, we need to qualify its drivers, its ingredients and its purpose for each step. We can do this by characterizing the trajectories of tourism diversification, based on Boschma et al. (2017) framework, which provides a theory of regional diversification. This framework mixes EEG (Boschma & Martin, 2007) and Transition Studies (Geels & Schot, 2007; Köhler et al., 2019) to overcome the limitations of each approach. It aims to qualify the place and path dependent character of diversification, which we try to apply to mountain tourism localized economies. Diversification can take different forms, according to its spatial and sectoral origins, and can then be related or unrelated, referring to the concept of related and unrelated variety in economic geography (Frenken et al., 2007). Regional diversification - and we assume tourism diversification - is related and place-dependent when the knowledge, resources and actors mobilized to develop alternative activities are those localized in the area of influence of the dominant economy. In our case, tourism diversification is related when its resources and actors are those already at the helm of the ski area. On the contrary, diversification is unrelated when the knowledge, resources and actors mobilized to develop alternative activities come from outside the localized economy, or in our case, outside the ski area. This distinction has never been applied to studies on the evolution of tourist destinations, but it can certainly complement the numerous works on tourism as an evolutionary science (Brouder et al., 2016; Gill, 2018).

Then, to propose a theory of regional diversification, Boschma et al. (2017) borrowed the concept of niche and regime (Geels & Schot, 2007) and used some insights from the field of transition studies, which further clarified the analysis of path dependence. In our case, considering ski-oriented tourism as the main regime in mid-mountain territories, diversification can be implemented by activating institutions, resources and technologies from the regime or from niches, driven by resources, institutions and actors that have never been involved in ski tourism. In addition to these criteria, Boschma et al. (2017) are interested in the notion of human agency - or 'bricolage and institutional entrepreneurship'- brought by Transition Studies. Human agency sheds light on the role of leading actors, institutions, and forms of actors organization, in blocking or facilitating transitions within socio-technical regimes. To be even more precise, we also follow several works on proximity studies and territorial governance (Chia et al., 2008; Gilly & Lung, 2005), which have already highlighted the role of human agency in the implementation of tourism diversification in mid-mountain areas structured by ski tourism (Achin, 2015).

By crossing the niche/regime character with the related/unrelated character of diversification, Boschma et al. (2017) define four theoretical archetypes of trajectories - from the most conservative to the most transformative - that we apply to tourism diversification: replication, transplantation, exaptation and saltation (Table 1).

**Table 1** – Adapting Boschma et al. (2017) theory to tourism diversification

| Path dependent | Place dependent | |
|---|---|---|
| | *Related* | *Unrelated* |
| *Regime* | **Replication** Development of activities complementary to skiing, requiring the same institutions as ski tourism | **Transplantation** Development of completely new activities, requiring the adaptation of local knowledge and institutions |
| *Niche* | **Exaptation** Development of new uses and new outlets for ski lift technologies or ski related product | **Saltation** Mutation of touristic activities and technologies, developing a completely new tourism system |

With this framework, our aim is to show the type of trajectory to which the observed dynamics of tourism diversification are linked. Then, we try to qualify the level of transformation of the mid-mountain areas structured by ski tourism by describing how these transformations take place, in which steps, with which tools and reconfigurations of stakeholders. Finally, we have taken the analysis further to correspond to studies on the evolution of ski tourism, by revealing the purpose of the tourism diversification trajectory. Indeed, tourism diversification can either promote local resources and anchor economic activities to a specific place, in a specification logic, or lead to the adoption of a production system dictated from the outside, reinforcing the affiliation to an economic sector rather than a territory, in a normalization logic (François et al., 2013). Among the different types of trajectories identified in the Boschma et al. (2017) framework, the distinction between normalization or specification is also useful to anchor the analysis of diversification and its role in the transformation of localized tourism economies.

**Material and method**

*Research method*

We used a qualitative method (Patton, 2014), grounded in our two study areas. Data collection consisted of (i) collecting historical information on the development of tourism in both study areas in historical books, scientific literature, and local newspaper archives in order to retrace the history of tourism diversification; (ii) collecting local administrations planning documents on tourism decisions and minutes of community and municipal councils; (iii) conducting face-to-face semi-structured interviews with stakeholders in the tourism diversification process, following the methodology of Beaud and Weber (1997).

We interviewed 61 tourism diversification stakeholders, some of them several times, for a total of 65 interviews. We wanted to cover the historical key actors in ski tourism as well as new actors in tourism activities to better understand the evolution inside and outside a tourism system around ski areas. The research sample for interviews is described in Table 2.

**Table 2** - Details of the research sample for interviews
The interview campaign lasted from January 2019 to July 2021

|  | *massif du Sancy* | *Haut-Chablais* |
|---|---|---|
| Number of interviewees<br>Of which: | **35**<br>4 local representatives<br>3 researchers<br>The head of the destination management organization<br>4 local administration technicians<br>3 heads of ski lifts companies<br>17 providers of tourist activities<br>2 farmers<br>1 public policy maker | **26**<br>8 local representatives<br>3 heads of destination management organizations<br>4 local administration technicians<br>1 head of a ski lifts company<br>1 technician of the Unesco Geopark<br>2 heads of a local associations<br>7 providers of tourist activities |
| In both study sites we participate to several collective meetings organized by public policy makers. We mainly met them on these occasions. | | |

We started with five exploratory interviews with local representatives, local administrations technicians and researchers to better understand the problem and to design our sampling strategy. Following a snowball logic (Corbin & Strauss, 2008), we continued with interviews with technicians in local administrations and tourist offices, local representatives, public policy makers, members of associations, providers of tourist activities, farmers and ski lift managers. The interviewees were asked about the strengths and weaknesses of their territory, the type of tourism activities they see coming and when it has arrived, the impact of these new activities, the leaders they have identified, the links they have with other stakeholders in tourism diversification, the learning needed to diversify, their needs in terms of adapting to global changes and their views on the future of their activity. 41 of these interviews were fully recorded and transcribed, others were just sources of historical information with notes. We then coded these notes and transcriptions thanks to the qualitative data analysis software N'Vivo (Woolf & Silver, 2017). After a pre-analysis, these codes were useful to identify themes and explanations about evolutionary mechanisms and reconfigurations related to tourism diversification at the territorial level.

After a preliminary historical analysis – based on historical information from the interviews, scientific papers and other local newspapers archives – we materialized tourism diversification trajectories, its events and periods, in a chronosystemic timeline (Bruley et al., 2021) for each study site, in order to see the evolution of tourism activities, tourism supply actors, and the spatial perimeter of tourism diversification in a specific territorial context. We read these timelines to identify the links between events and elements of diversification, which constitute a specific configuration. Finally, we described changes in these configurations to explain bifurcations in the trajectory, their drivers and directions. During their elaboration, we presented these chronosystemic timelines several times to the stakeholders of tourism diversification for each study site. These exchanges enriched and confirmed their elaboration, challenging and confronting the Boschma et al. (2017) framework with the observed dynamics, in an abductive and grounded approach (Boudon & Bourricaud, 1989; Glaser & Strauss, 1967).

*Study sites*

Our approach is based on two study sites, which are the perimeter of the massif du Sancy in the Massif central, and the Haut-Chablais in the French Northern Alps. The two study sites are located in the Auvergne-Rhône-Alpes Region of France, which has 75% of its territory classified as mountainous and being the cradle of alpine skiing in France. In this region, in the 1960s and 1970s, the ski tourism industry was the main basis for the development of peripheral mid-mountain areas. To take into account territorial dynamics and to illustrate the notion of *territoires* (Pachoud et al., 2022) in the French geography literature, we focused on the perimeter of intercommunalities where several ski resorts are located (figure 1 and 2).

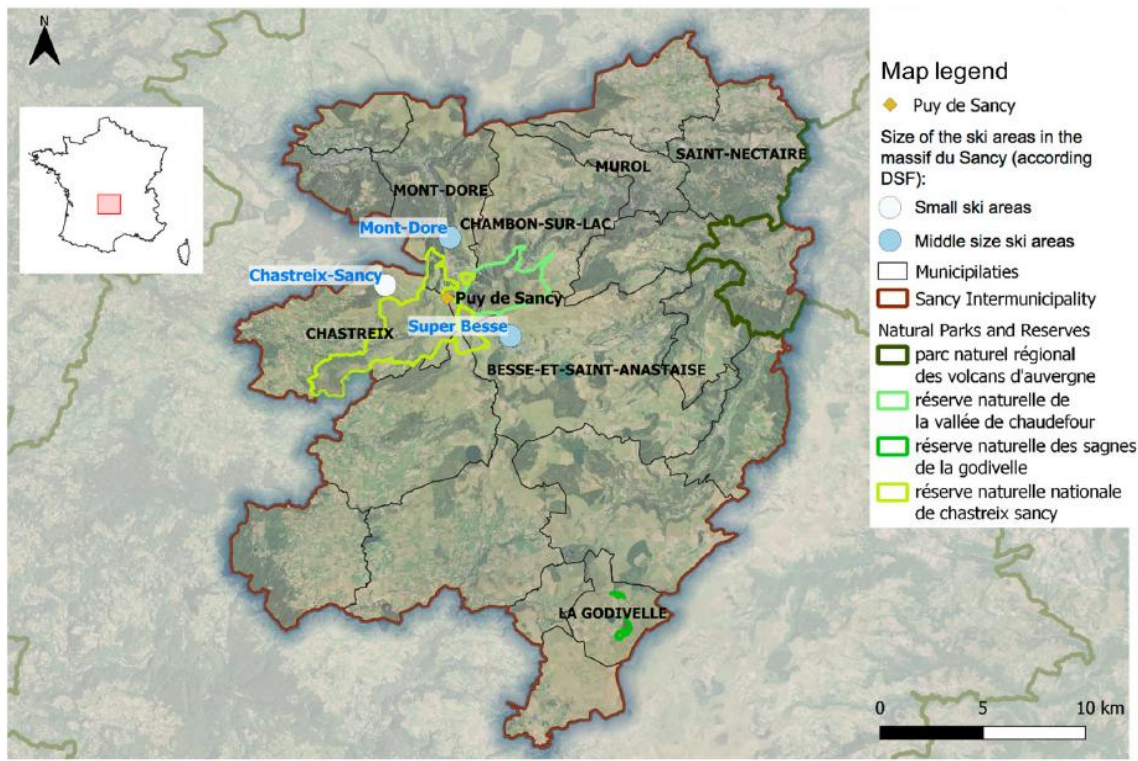

*Figure 1 - Ski resorts, natural parks and reserves spatial organization on the massif du Sancy perimeter. The three ski areas are located between 1070 and 1880 m asl. The classification of ski resorts is explained in Table 3.*

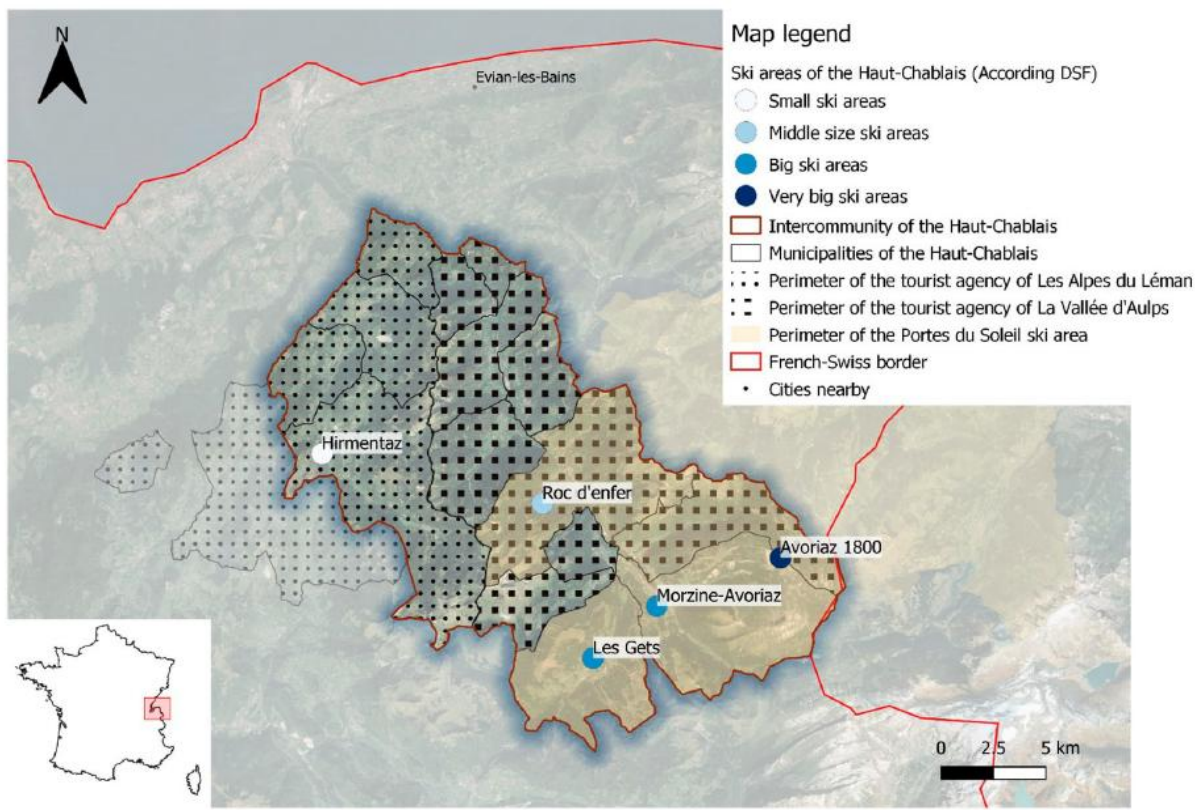

*Figure 2 - Ski resorts and tourism planning system spatial organization on the Haut-Chablais perimeter. The five ski areas are located between 945 and 2260 m asl.*

In both study sites, skiing is not the only tourist product offered and alternatives, such as outdoor sports - trail, downhill bike, etc. - cultural tours, agritourism or geological tourism in the Haut-Chablais, have been implemented for more than 30 years now. In both cases, the development of these numerous alternative products is supported by public policies, such as the 'Pôle de pleine nature' program in the massif du Sancy - which supports the development of all-year-round tourism in order to increase the residential attractiveness - or the 'Espaces Valléens' program in the Haut-Chablais - which supports the development of a four-season and sustainable tourism in order to promote adaptation, transition and a diversified economic development. It is important to note here that the tourism diversification in each French massif is directly linked to public policies, with supra-local actors driving the implementation of tourism diversification in France (Achin, 2015). The main difference between the two cases is expressed in global territorial features, such as the demography, but also in the ski resorts characteristics – detailed in Table 3 – and in the number of tourist overnight stays and in the tourist seasons. The massif du Sancy, struggle since the 1960s with a strong decline in the local population, stabilizing around 9641 permanent inhabitants in 2019. It is therefore the main tourist destination of the Massif Central, registering 2,4 million overnight stays in 2021, with a similar distribution between summer and winter. In the Haut-Chablais, the dynamic is the opposite, with an increase in the local population and reaching 12821 inhabitants in 2019. The perimeter of the Portes du Soleil, which is slightly larger than the perimeter of the Haut-Chablais intercommunity, register 3.7 million overnight stays in 2021. However, before the Covid-19 crisis, it registered 7 million overnight stays in 2019, of which 64% were realized in winter and 34% in summer. The reasons for these differences are described in the following section.

**Table 3**: Main characteristics of the different ski resorts located in the two study sites in 2022

| Ski resorts | Type of management | Number of ski lifts | Number of ski slopes | Capacity of the ski lifts (pers/km/h)* |
|---|---|---|---|---|
| *Massif du Sancy:* | | | | |
| - Super Besse | Semi-public | 21 | 27 | 4950 |
| - Mont Dore | Semi-public | 14 | 31 | 2667 |
| - Chastreix-Sancy | Semi-public | 7 | 18 | 508 |
| *Haut-Chablais:* | | | | |
| - Avoriaz 1800 | Private | 27 | 50 | 18826 |
| - Morzine-Avoriaz | Private | 48** | 69 | 9204 |
| - Les Gets | Semi-public | | 71 | 10489 |
| - Le Roc d'Enfer | Semi-public | 16 | 32 | 3100 |
| - Bellevaux-Hirmentaz | Semi-public | 15 | 25 | 2115 |

*\* The capacity of ski lift, expressed in number of persons per kilometer traveled by the ski lift, is used to compare the volume of the fleet between different ski lift operators. DSF (Domaines Skiables de France), the main syndicate for ski lift operators, uses this indicator to classify French ski resorts into four categories: Very big ski resorts, big ski resorts, medium size ski resorts and small ski resorts. These categories for the ski resorts in our study areas are illustrated by the Figure 1.*
*\*\* The Morzine-Avoriaz and Les Gets ski resorts have the same ski area, they share the same ski pass.*

**Results**

Our analysis of tourism diversification from an evolutionary perspective has yielded two main results. First, tourism diversification is a specific trajectory in ski-tourism oriented *territoires,* following three steps in a non-linear manner and linked to the specificities of each territory. We characterize tourism diversification trajectories following a replicable framework for other destinations. Second, by applying the framework of Boschma et al. (2017) to winter tourism destinations, it appears that tourism diversification trajectories can be of different nature, leading to different degrees of transformation of the local economy. This transformation leads to the integration of ski resorts and a snow-based economic model, within a sectoral logic, into a broader and more diversified economy at a territorial scale.

*The three steps in tourism diversification trajectories in mid-mountain territories*

By analyzing the changes in sequences in the chronosystemic timelines, we identified three main steps in tourism diversification trajectories that are common to both study sites. The following three steps are chronological: (1) the way tourism is established, (2) the development of alternative tourism activities to skiing, (3) the structuring of the diversified tourism offer. For the corresponding steps, we identify the path and place dependent nature, using the criteria of Boschma et al. (2017), as well as the way in which the local resources and know-how are promoted.

*The way tourism becomes a pillar of the local economy*

In the light of the evolution of the two study areas, the tourism diversification trajectories are defined by the way tourism becomes a pillar of the local economy. Thanks to historical and local communities documents, we can affirm that, in the massif du Sancy, tourism began with thermalism, in the 1830s, which led to the construction of tourist infrastructures. Later, ski infrastructures were built in 1936, creating the Mont-Dore ski resort, then the Super Besse ski resort was opened in 1961, followed by two smaller ones in 1969. In the Haut-Chablais, tourism began with a few summer stays and health centers, but in the 1930s, ski tourism became the mainstay of the local economy. The ski area of Les Gets opened in 1934 and that of Morzine, sharing the same ski area, in 1937. After the Second World War, the economy of the Haut-Chablais specialized in the operation of ski lifts, with the great real estate project of Avoriaz, inaugurated in 1967, appealing to international customers. Ski areas then became central to the local economy, but also to the functioning of the territory, with local public and economic actors organizing and collaborating for the operation of ski lifts. This gave rise to a regime based on ski tourism.

*The development of alternatives to downhill ski*

After the establishment of ski tourism, the development of alternative activities to skiing has been observed in each of the study areas since the 1970s. In the Massif du Sancy, it began with a diversification of actors, with public actors participating in the planning of tourism. In 1972, the municipalities with ski resorts and their neighbors formed a syndicate to help them promote all tourism, not just winter tourism. This syndicate, became the basis for local planning and territorial development around year-round tourism, and led to the creation of an intercommunality around tourism in 2000. Public actors became very important in tourism - because 'municipalities and their mayors are ceo of semi-public tourism companies' as an employee of the tourist office explained - and this intercommunality was then 'a trigger in the sense that it forced them to sit and ask questions together' according an economic actor. Since then, these public actors have also offered new tourist activities, such as the creation of a land art festival in 2007. In the 1980s, new economic actors settled in Sancy as mountain guides, or with the purpose of offering some activities of discovery of the local heritage, or fishing, among others. Finally, in the 2010s, the ski lift operator of Super Besse followed the trend and started to diversify its activities with downhill biking, escape games or zip lines, opening in summer. These diversification initiatives ultimately follow a common diversification dynamic at the scale of the massif (Rouch, 2022). In the Haut-Chablais, the deployment of alternatives to skiing began on the ski resort scale. In 1973, the initiator of the Avoriaz project launched a fantastic film festival that lasted until 1993. In the 1990's, cultural activities gave way to downhill mountain biking, using ski lifts in the summer. This practice has grown to such an extent that the resort of Les Gets has regularly hosted international championships since the 1990s. It has created a bi-seasonal tourism system on the scale of international ski resorts, leading to a first diversification path driven by the operation of ski lifts. But a summer season for the ski lifts, which, according to a local representative 'only represents the turnover of a weekend in February', justifying the little credit given to alternatives. At the same time, as in the massif du Sancy, the Haut-Chablais and its environs have seen the settlement of economic actors offering outdoor activities since the 1980s. The public authorities have also positioned themselves on this diversification axis by promoting local natural, cultural and geological sites, notably with the designation of the Chablais Unesco Geoparc in 2012. They are also promoting local know-how, such as agritourism and craftsmanship, by creating a terroir and know-how trail in 2017. Both on the outskirts of the resorts and on their periphery, these initiatives represent a second wave of diversification. Two parallel paths of tourism diversification can thus be observed and coexist in Haut-Chablais. This situation is quite complex, especially since the local tourism organization is particularly complicated, with 5 tourist offices within the perimeter of Haut-Chablais.

*Structuring tourism diversification*

This situation reveals a third stage of structuring this diversified offer, which is quite complex in Haut-Chablais. The two tourism diversification paths and the complex organization confirm the existence of three diversification subspaces. The international perimeter of ski resorts is the support for the first diversification path. Then, the second diversification path is supported by two sub-perimeters, known as the lower part of the Aulps valley and the Brevon valley, which retain distinct territorial features. In spite of the efforts to group together in an inter-municipality and the numerous tourism diversification initiatives driven by a variety of stakeholders, there is no global structuring of tourism diversification at the territorial level (see the multiple perimeter of action in figure 1). On the other hand, in the Massif du Sancy, we can observe the structuring of the diversified tourism offer at the massif scale. However, as one economic actor explains, 'we had to fight to create this thing. [...] You told the shopkeepers of Le Mont-Dore that they were going to be with those of Super Besse, and vice versa, it was just not possible'. This structuring was a long and difficult work, driven by the inter-community tourist office, leader in the evolution of tourism.

***Which forms of tourism diversification for which territorial transformations?***

From the analysis of these three stages and with this long-term perspective, we can now study and reveal the transformative nature of tourism diversification trajectories. To do so, we focus on the origin of the actors, the knowledge and the capacities involved in these stages along the diversification trajectory and on the degree of valorization of local specificity followed in the diversification of tourism offer. We then determine to which category the tourism diversification trajectories of the massif du Sancy and the Haut-Chablais belong to, between replication, transplantation, exaptation or saltation (Boschma et al., 2017). The category is then an indicator to explain the extent to which tourism diversification transforms the local economy.

In the massif du Sancy, the first stage of the tourism diversification trajectory reveals the existence of actors, knowledge and resources in thermalism that existed before the specialization in ski tourism in the 1960s. Thermalism and the inclusion of ski tourism in a global offer acted as a fertile ground for the development of alternatives. In the second step, the actors, activities and knowledge mobilized in the tourism diversification movement come mostly from niches, as they are not directly linked to the operation of ski resorts. They are also unrelated because they come from outside the spatial perimeter of the Sancy massif and mobilize capacities on the outskirts of ski resorts. In fact, since the 1970s, the municipalities involved in tourism planning have been those that support the ski resorts, but also those on the periphery of the slopes of the Puy de Sancy. It is the same for private providers of tourist activities, with several located on the perimeter of ski resorts, but most of them located throughout the northern part of the massif (Achin, 2015). All these tourism diversification products and initiatives form a common tourism diversification trajectory at the territorial level, following a saltation process (table 4). A saltation driven by local representatives, public policies and by the local intercommunity tourist office - or destination management organization - whose main mission was, and still is, to get out of a ski tourism regime scale and promote territorial development. It is even clearer as most of the tourism diversification initiatives try to promote the local landscape, culture and heritage.

**Table 4** - The different forms of tourism diversification trajectory in the massif du Sancy and the Haut-Chablais depending on their origin and nature

| Territoire | Niche / regime (*path dependent*) | related / unrelated (*place dependent*) | Valorization of local specificities |
|---|---|---|---|
| **Massif du Sancy** | Mainly niche | Unrelated | Partial |
| | **Saltation** | | |
| **Haut-Chablais** | Ski areas : From niches, then regime | Related | In a logic of amusement park |
| | **Replication and transplantation** | | |
| | In the rest of the territoire : Niche | A majority of unrelated | Valorization of local resources and know-how |
| | **Saltation** | | |

However, the situation is not so neat, as we observe a mix of diversification processes within ski areas. In the words of Boschma et al., (2017), we see processes of replication - with the development of new winter products such as freestyle snow parks - but also processes of transplantation and exaptation - with the development of zip lines, a product that exists in other types of regimes and offers a new application for the technologies of ski-tourism regimes. The processes of replication and transplantation continue the actual ski-tourism regime, and the exaptation implies a transformation of the knowledge needed to operate the tourism system. The Super-Besse ski area is a great example of articulation of the three processes, where we observe permanency of a regime around ski-tourism along attempts of knowledge transformation. In conclusion, ski resorts are not isolated from the rest of the territory and are included in a more global and diversified tourism system. Even if we see different types of diversification processes at a more local level, they are included, in a coherent way, in a territory-wide diversification trajectory following saltation.

Tourism diversification trajectories of the Haut-Chablais began later than in the massif du Sancy and in a much more specialized context, considering the three internationally renowned ski resorts. Moreover, since the 1970s, marked by the second step in the implementation of diversification, we have noted the coexistence of two different trajectories. The first, at the level of the two municipalities supporting the internationally renowned ski resorts, is a conservative trajectory of tourism diversification. The actors, means and resources mobilized in the implementation of the Avoriaz Film Festival and the practice of downhill biking are largely the same as those of the regime around the ski resorts. Even if the tourist offer has evolved and new activities have been proposed, the network of stakeholders at this scale hasn't changed since the 1960s, is still closed, dominates the evolution of tourism and is not open to other actors of tourism diversification (Rouch et al., 2022). The ski lift operator, the municipality that supports the ski area, the tourist offices, the Association of Local Ski Areas of Les Portes du Soleil, real estate investors, ski related shops, ski related activity providers and host companies are well structured around ski lift operations, whether for winter skiing or summer downhill biking. Here, the drivers of tourism diversification are related to existing local strengths, such as those used to enhance the stay of tourists during their ski week. With the presence of British residents in Haut-Chablais since the 2010s, we also see a diversification in the local offer, but which stays in a ski regime perspective. One of the mayors explains here that 'they invent new services, new modes of operation to attract and offer ski stays, like all-in stays, which disrupt the initial offer'. This diversification trajectory of tourism therefore mostly follows processes of replication and transplantation. With the development of downhill mountain biking, we note the existence of an initial logic of exaptation, since the practice of mountain biking was a niche in the 1990s and allowed ski lift operators to find new opportunities for existing technology. But they are not integrated on a territorial scale and remain in a sector and company dynamic. Processes of transplantation, replication and a bit of exaptation do not transform and question the contours of the ski regime. Rather than perpetuating a regime around skiing, some of the main actors of this trajectory see tourism diversification as a way to perpetuate mass tourism in the mountains. The development of generic tourist products and infrastructures, such as a summer toboggan run or a park of lights in the trees in the ski resort of Les Gets, is more likely to be envisaged by local tourist activity providers in order to create a typical 'amusement park' offer - as stated by a shopkeeper - rather than to promote local specificities. Table 4 summarizes this analysis and allows us to situate this type of conservative tourism diversification trajectory in relation to the second, which we have seen emerge on a more global scale in Haut-Chablais.

This second trajectory is driven by a variety of private and public initiatives, on a more territorial scale, and was initially triggered by the development and affirmation of new practices such as rafting in the 1980s. The providers of tourist activities and the intermunicipal administrations, that have begun to structure these new products, have mobilized institutions, means and methods of operation that have never been observed in the trajectory of the system around the international ski resorts and that are mostly unrelated. In fact, even if some of these actors are located on the perimeter of the international ski resorts, the majority of them are located on their periphery and in the rest of the territory. This difference in spatial scale is also felt in terms of the reference perimeter of diversification expressed by the actors. Compared to the actors of the first trajectory, the actors of the second trajectory have a wider perimeter of action. They include themselves as actors of the different valleys of the Chablais region, not only of the Haut-Chablais, with a tourist activity provider feeling like a 'people in the middle' of several different spaces. Whereas the actors of the first trajectory hardly transcend the scale of the commune, as illustrated by this mayor: 'Chablais, Chablais, I only know Morzine'. In addition, most of the actors of this second trajectory set up their tourism projects in order to appreciate the specificities of the territory, such as the specific and remarkable environment or the geological heritage of Haut-Chablais. They also seek to be anchored in a logic of promoting local knowledge and craftsmanship. This diversification path thus takes the form of saltation (Table 4), but finds less resonance with local elected officials and decision-makers, as its networks of stakeholders are less structured than those of the ski regime (Rouch et al., 2022).

**Discussion and conclusion**

Since the study of tourism diversification as a climate change adaptation strategy obscures its temporal and spatial dimensions, in this paper we focused on the process itself in order to understand its possible impacts at the territorial scale in the long term. Our study highlights the relevance of the trajectory framework proposed by Boschma et al. (2017) to reveal the transformative effects of tourism diversification. The mid-mountain areas, specialized in the winter sports tourism sector but trying to diversify the tourism offer for more than 30 years now, were excellent study areas to apply this framework. Subscribing to both transition studies and EEG, we used this framework and added a territorial approach to understand the evolution of tourism in mid-mountain territories structured by ski tourism. Despite the inherent limitations of our qualitative and grounded approach – such as the cumbersome of data collection limiting to only two study sites, or our attempt to combine different types of data without necessarily spelling out how they complement each other – our study brings up three main points for discussion and outlooks.

First, we confirm the existence of the different types of trajectories identified by Boschma et al. (2017). In both study areas we see similar diversification processes, forming a single tourism diversification trajectory in the massif du Sancy, but two different trajectories in the Haut-Chablais. In the massif du Sancy, the diversification trajectory takes the form of a saltation, qualified as a 'complete regime change', requiring a 'transformation of regional capacities' (Boschma et al., 2017), which then leads to the establishment of a diversified tourism system (Rouch, 2022) around a specification dynamic (François et al., 2013). In contrast, the case of the Haut-Chablais shows a limitation in the vision of Boschma et al. (2017). While the authors focus on the regional scale and assume that a single diversification trajectory can take place in an economic region, we distinguish two trajectories that coexist at the same spatial scale but are driven by different actors. One in the form of replication and transplantation at the scale of ski areas - more conservative, in the continuity of industrialized ski tourism - and another in the form of saltation, less structured, but in a logic of territorial development at the scale of Haut-Chablais. This result is an expression of a co-evolution (Ma & Hassink, 2013) and calls into question the interweaving of scales that supports the diversification of tourism and leads to overcome a generic notion of diversification as one unit. In the massif du Sancy, the ski area is integrated in a more territorial scale, whereas in the Haut-Chablais, the two trajectories divide the intercommunal area into several perimeters of action that are not aligned on a global scale, despite the complementarity of the offers. Thus, the type of trajectory is important, but so is the scale in which it takes place. In addition, with respect to our study cases, scale is important, but so is time. Boschma et al. (2017) did not contextualize the four different types of diversification trajectories in time, so we could consider these categories like steps in diversification trajectories. To test this, more studies on other mid-mountain territories could apply Boschma et al., (2017) framework to better understand and potentially see the emergence of invariants or peculiarities in a series of tourism and regional diversification trajectories.

Second, the consideration of time challenges the concept of breaks or tipping points as used in Transitions Studies. According to our analysis, the diversification of tourism can become a break in terms of actors, economic activities and spatial perimeter of a regime around ski tourism. However, in the case of the massif du Sancy, this break does not mean the end of the previous regime around skiing. It seems to be a construction that integrates ski tourism into a broader tourism system. Moreover, the diversification of tourism is more incremental than a brutal temporal shift. We point out here that tourism diversification may not be a path creation (Clavé & Wilson, 2017; Gill, 2018), but rather a synonym of path plasticity, known as an incremental path out of dependence (Halkier & Therkelsen, 2013). According to Halkier and Therkelsen (2013), path plasticity at the scale of tourist destinations is achieved through two processes: the ability to offer and organize year-round activities, and the sharing of knowledge and know-how between different levels of actors. In this paper, we have focused on the first process and our current approach is quite limited to conclude that tourism diversification is a path plasticity, but further works have shown that tourism diversification can also promote learning and knowledge sharing (Rouch, 2022). This discussion concerning the notion of breaks is all the more lively given that our trajectory approach tend to smooth changes depending on the time scale of analysis. It is one main limitation we did not address in this paper about the degree of radicalism and importance of the changes associated with tourism diversification at one specific time in the trajectory. Again, time should be more closely monitored in tourism or regional diversification studies.

Finally, as Boschma et al. (2017) wanted to address by calling for Transition Studies, actors and the way they organize themselves in networks are the drivers of transformations. We explained how, thanks to a territorial approach that puts actors and their dynamics at the center (Gumuchian et al., 2003). We could explain these transformations even more precisely by extending the analysis to other territories and more study sites, but also by describing how some actors can block change by sacralizing lock-in effects (Berard-Chenu et al., 2022) or, on the contrary, be innovation drivers in defining future trajectories. Here, as a research outlook, a renewed reading of the evolution of mid-mountain tourist destinations with a sociology of innovation perspective could be useful (Callon, 1986; Paget et al., 2010).

**Declaration of interest**

No potential conflict of interest was reported by the author(s).

# References

Abegg, B., Agrawala, S., Crick, F., & De Montfalcon, A. (2007). Climate change impacts and adaptation in winter tourism. In [w:] S. Agrawala (red.) (Ed.), *Climate Change in the European Alps: Adapting Winter Tourism and Natural Hazards Management*. OECD Publishing, Paris.

Achin, C. (2015). *La gouvernance de la diversification comme enjeu de l'adaptation des stations de moyenne montagne: l'analyse des stations de la Bresse, du Dévoluy et du Sancy [The governance of diversification as a challenge for the adaptation of mid-mountain resorts: analysis of the resorts of Bresse, Dévoluy and Sancy]*, Université Grenoble Alpes. https://tel.archives-ouvertes.fr/tel-02603442v2

Beaud, S., & Weber, F. (1997). Guidelines for field surveys. Producing and analysing ethnographic data. *La Découverte, Paris*.

Benur, A. M., & Bramwell, B. (2015). Tourism product development and product diversification in destinations. *Tourism management*, *50*, 213–224. https://doi.org/10.1016/j.tourman.2015.02.005

Berard-Chenu, L., François, H., Morin, S., & George, E. (2022). The deployment of snowmaking in the French ski tourism industry: a path development approach. *Current Issues in Tourism*, 1–18. https://doi.org/10.1080/13683500.2022.2151876

Boschma, R., Coenen, L., Frenken, K., & Truffer, B. (2017). Towards a theory of regional diversification: combining insights from Evolutionary Economic Geography and Transition Studies. *Regional Studies*, *51*(1), 31–45. https://doi.org/10.1080/00343404.2016.1258460

Boschma, R., & Martin, R. (2007). Constructing an evolutionary economic geography. *Journal of Economic Geography*, *7*(5), 537–548. https://doi.org/10.1093/jeg/lbm021

Botti, L., Goncalves, O., & Peypoch, N. (2012). Analyse comparative des destinations «neige» pyrénéennes [Comparative analysis of Pyrenean snow destinations]. *Journal of Alpine Research/ Revue de géographie alpine*(100-4). https://doi.org/https://doi.org/10.4000/rga.1843

Boudon, R., & Bourricaud, F. (1989). *A critical dictionary of sociology*. University of Chicago Press.

Bramwell, B. (2004). *Coastalmass tourism: Diversification and sustainable development in Southern Europe*. Channel View Publications.

Brouder, P., Clavé, S. A., Gill, A., & Ioannides, D. (2016). Why is tourism not an evolutionary science? Understanding the past, present and future of destination evolution, *Tourism Destination Evolution* (pp. 13–30). Routledge. https://doi.org/10.4324/9781315550749-7

Brouder, P., et Eriksson, R. H. (2013). Tourism evolution: On the synergies of tourism studies and evolutionary economic geography, *Annals of Tourism Research*,43, pp. 370-389.  doi: https://doi.org/10.1016/j.annals.2013.07.001

Bruley, E., Locatelli, B., Vendel, F., Bergeret, A., Elleaume, N., Grosinger, J., & Lavorel, S. (2021). Historical reconfigurations of a social–ecological system adapting to economic, policy and climate changes in the French Alps. *Regional Environnemental Change*, *21*(2), 34–15. https://doi.org/10.1007/s10113-021-01760-8

Burki, R., Elsasser, H., & Abegg, B. (2003). *Climate change and winter sports: environmental and economic threats*. University of Zurich. https://raonline.ch/pages/edu/pdf5/burkirep01a.pdf

Butler, R. W. (1980). The concept of a tourist area cycle of evolution: implications for management of resources. *Canadian Geographer/Le Géographe canadien*, *24*(1), 5-12. https://doi.org/10.1111/j.1541-0064.1980.tb00970.x

Butler, R. W. (2006). *The tourism area life cycle: Conceptual and theoretical issues*  (Vol. 2). Channel View Publications.

Callon, M. (1986). Some elements of a sociology of translation: domestication of the scallops and the fishermen of St Brieuc Bay. *The sociological review*, *32*, 196-233.

Capone, F. (2006). Systemic approaches for the analysis of tourism destination: towards the tourist local systems. In Routledge (Ed.), *Tourism local systems and networking* (pp. 7-23). https://www.taylorfrancis.com/chapters/edit/10.4324/9780080462387-10/systemic-approaches-analysis-tourism-destination-towards-tourist-local-systems-francesco-capone

Chabault, D. (2009). *Gouvernance et trajectoire des réseaux territoriaux d'organisations: une application aux pôles de compétitivité [Governance and trajectory of territorial networks of organizations: an application to competitiveness clusters]*, Université de Tours.

Chia, E., Torre, A., & Rey-Valette, H. (2008). Conclusion: Vers une «technologie» de la gouvernance territoriale! Plaidoyer pour un programme de recherche sur les instruments et dispositifs de la gouvernance des territoires [Conclusion: Towards a "technology" of territorial governance! Plea for a research program on the instruments and devices of territorial governance]. *Norois*, *209*(4), 167-177. https://doi.org/https://doi.org/10.4000/norois.2603

Clavé, S. A., & Wilson, J. (2017). The evolution of coastal tourism destinations: A path plasticity perspective on tourism urbanisation. *Journal of Sustainable Tourism*, *25*(1), 96-112. https://doi.org/https://doi.org/10.1080/09669582.2016.1177063

Corbin, J., & Strauss, A. (2008). Strategies for qualitative data analysis. *Basics of Qualitative Research. Techniques and procedures for developing grounded theory*, *3*(10).

Erkuş-Öztürk, H., & Terhorst, P. (2015). Economic diversification of a single-asset tourism city: evidence from Antalya. *Current Issues in Tourism*, *21*(4), 422–439. https://doi.org/10.1080/13683500.2015.1091806

Farmaki, A. (2012). A supply-side evaluation of coastal tourism diversification: the case of Cyprus. *Tourism Planning & Development*, *9*(2), 183-203. https://doi.org/https://doi.org/10.1080/21568316.2011.634431

François, H. (2007). *De la station ressource pour le territoire au territoire ressource pour la station: le cas des stations de moyenne montagne périurbaines de Grenoble [From the resort as a resource for the territory to the territory as a resource for the resort: the case of the peri-urban mid-mountain resorts of Grenoble]*, Univsersité Grenoble 1. https://tel.archives-ouvertes.fr/tel-00185781

François, H., Hirczak, M., & Senil, N. (2013). De la ressource à la trajectoire: quelles stratégies de développement territorial? [From resource to trajectory: which territorial development strategies?], *Géographie, économie, société*, *15*(3), 267-284. https://www.cairn.info/revue-geographie-economie-societe-2013-3-page-267.htm?contenu=resume

Frenken, K., Van Oort, F., & Verburg, T. (2007). Related Variety, Unrelated Variety and Regional Economic Growth. *Regional Studies*, *41*(5), 685–697. https://doi.org/10.1080/00343400601120296

Geels, F. W., & Schot, J. (2007). Typology of sociotechnical transition pathways. *Research policy*, *36*(3), 399-417. https://doi.org/https://doi.org/10.1016/j.respol.2007.01.003

George, E., & Achin, C. (2020). Implementation of tourism diversification in ski resorts in the French Alps: a history of territorializing tourism, *Local Resources, Territorial Development and Well-being*. Edward Elgar Publishing. https://doi.org/10.4337/9781789908619.00014

Gill, A. M. (2018). Challenges to the resilience of Whistler's journey towards sustainability, *Tourism in transitions* (pp. 21-37). Springer. https://doi.org/10.1007/978-3-319-64325-0_2

Gill, A. M., et Williams, P. W. (2014). Mindful deviation in creating a governance path towards sustainability in resort destinations, *Tourism Geographies*, 16(4), pp. 546-562. doi: 10.1080/14616688.2014.925964

Gilly, J.-P., & Lung, Y. (2005). Proximities, industries and territories (In French). *Cahiers du GRES (2002-2009)*. https://ideas.repec.org/p/grs/wpegrs/2005-09.html

Glaser, B., & Strauss, F. (1967). *The discovery of grounded theory : strategies for qualitative research*. (1967)

Gumuchian, H., Grasset, E., Lajarge, R., & Roux, E. (2003). *Les acteurs, ces oubliés du territoire*. Economica.

Halkier, H., & Therkelsen, A. (2013). Exploring tourism destination path plasticity. The case of coastal tourism in North Jutland, Denmark. *Zeitschrift für Wirtschaftsgeographie*, *57*(1-2), 39–51. https://doi.org/10.1515/zfw.2013.0004

Hjalager, A.-M. (1996). Agricultural diversification into tourism: Evidence of a European Community development programme, *Tourism management*, 17(2), pp. 103-111. doi: https://doi.org/10.1016/0261-5177(95)00113-1

Ioannides, D., Halkier, H. and Lew, A. (2015). Evolutionary economic geography and the economies of tourism destinations, *Tourism Geographies* 16(4), 535-9.

Köhler, J., Geels, F. W., Kern, F., Markard, J., Onsongo, E., Wieczorek, A., Alkemade, F., Avelino, F., Bergek, A., & Boons, F. (2019). An agenda for sustainability transitions research: State of the art and future directions. *Environmental innovation and societal transitions*, *31*, 1-32. https://doi.org/https://doi.org/10.1016/j.eist.2019.01.004

Langenbach, M., & Jaccard, É. (2019). Innovation at the heart of tourism diversification in mountain resorts? A critical approach to the role of trail running in Switzerland. *Mondes du Tourisme*(15). https://journals.openedition.org/tourisme/1936

Ma, M., & Hassink, R. (2013). An evolutionary perspective on tourism area development. *Annals of tourism research*, *41*, 89-109.

Marcelpoil, E. (2008). *Les trajectoires d'évolution des destinations touristiques de montagne [The evolutionary trajectories of mountain tourist destinations]*, Université de Pau et des Pays de l'Adour. https://hal.inrae.fr/tel-02590601

Niewiadomski, P., & Brouder, P. (2022). Towards an evolutionary approach to sustainability transitions in tourism. In *Handbook of Innovation for Sustainable Tourism* (pp. 82-111). Edward Elgar Publishing.

Pachoud, C., Koop, K., & George, E. (2022). Societal transformation through the prism of the concept of territoire: A French contribution. *Environmental innovation and societal transitions*, *45*, 101–113. https://doi.org/10.1016/j.eist.2022.10.001

Paget, E., Dimanche, F., & Mounet, J.-P. (2010). A tourism innovation case: An Actor-Network Approach. *Annals of tourism research*, *37*(3), 828–847. https://doi.org/10.1016/j.annals.2010.02.004

Pascal, R. (1993). *Problèmes structurels des stations de moyenne montagne[Structural problems of mid-mountain resorts]*

Patton, M. Q. (2014). *Qualitative research & evaluation methods: Integrating theory and practice*. Sage publications.

Perret, J. (1992). *Le développement touristique local: les stations de sports d'hiver. [Local tourism development: winter sports resorts.]* https://hal.inrae.fr/tel-02575241

Perrin-Malterre, C. (2018). Tourism diversification process around trail running in the Pays of Allevard (Isère). *Journal of Sport & Tourism*, *22*(1), 67–82. https://doi.org/10.1080/14775085.2018.1432410

Rouch, L. (2022). *Tourism diversification trajectories in mid-mountain territories : stakeholder networks and learning dynamics* [in French], Université Grenoble Alpes. https://www.theses.fr/s226774

Rouch, L., George, E., & Rieutort, L. in press. Évolutions des réseaux d'acteurs le long de trajectoires de diversification touristique en moyenne montagne : de nouvelles formes de collaborations ? [Changes in stakeholder networks along tourism diversification trajectories in mid-mountain areas: new forms of collaboration?], *Géographie, économie et sociétés*.

Sanz-Ibáñez, C., et Clavé, S. A. (2014). The evolution of destinations: towards an evolutionary and relational economic geography approach, *Tourism Geographies*,16(4), pp. 563-579. doi: 10.1080/14616688.2014.925965

Sharpley, R., & Vass, A. (2006). Tourism, farming and diversification: An attitudinal study. *Tourism management*, *27*(5), 1040–1052. https://doi.org/10.1016/j.tourman.2005.10.025

Steiger, R., Scott, D., Abegg, B., Pons, M., & Aall, C. (2019). A critical review of climate change risk for ski tourism. *Current Issues in Tourism*, *22*(11), 1343–1379. https://doi.org/10.1080/13683500.2017.1410110

Weidenfeld, A. (2018). Tourism Diversification and Its Implications for Smart Specialisation. *Sustainability*, *10*(2), 319. https://doi.org/10.3390/su10020319

Woolf, N. H., & Silver, C. (2017). *Qualitative analysis using NVivo: The five-level QDA® method*. Routledge.